\documentclass[journal]{IEEEtran}
\usepackage{psfrag}
\usepackage{amsmath}
\usepackage{subfigure}
\usepackage{url}
\usepackage{stfloats}
\usepackage{cite}
\usepackage{amsmath}
\usepackage{mathtools}
\usepackage{upgreek}
\usepackage{amssymb}
\usepackage{amsfonts}
\usepackage{graphicx}
\usepackage{fancybox}
\usepackage{color}
\usepackage{algorithm}
\usepackage{algorithmic}
\usepackage{multirow}
\usepackage{booktabs}
\usepackage{threeparttable}
\usepackage{latexsym}
\usepackage{bm,array}
\usepackage{graphicx}
\usepackage{float}
\usepackage{hyperref}
\usepackage{acronym}
\usepackage{textcomp}

\IEEEoverridecommandlockouts
\normalsize

\begin{document}
\title{\textcolor{black}{QoS-Compliant 3D Deployment Optimization Strategy for UAV Base Stations} \vspace{0.2cm}}
\author{Xukai Zhong, Yiming Huo,~\IEEEmembership{Member,~IEEE}, Xiaodai Dong,~\IEEEmembership{Senior Member,~IEEE}, and Zhonghua Liang,~\IEEEmembership{Senior Member,~IEEE}
\thanks{X. Zhong, Y. Huo and X. Dong are with the Department of Electrical and Computer Engineering, University of Victoria, Victoria, BC V8P 5C2, Canada (e-mail: xukaiz@uvic.ca, ymhuo@uvic.ca, xdong@ece.uvic.ca). This work was supported by Wighton Engineering Product Development Fund. (\emph{Corresponding author: Xiaodai Dong})} 
 \thanks{Z. Liang is with School of Information Engineering, Chang’an University, Xi’an, Shanxi Province, China (e-mail: lzhxjd@hotmail.com). Z.  Liang’s work was supported in part by  the Natural Science Basic Research  Project in Shaanxi Province of China under Grant 2020JM-242, and in part by the National Natural Science Foundation of China under Grant 61871314, and in part by the Fundamental Research Funds for the Central Universities, CHD under Grant 300102249303.} 

\vspace{-0.4cm}} 

\maketitle

\begin{abstract}
Unmanned aerial vehicle (UAV) is being integrated as an active element in 5G and beyond networks. Because of its flexibility and mobility, UAV base stations (UAV-BSs) can be deployed according to the ground user distributions and their quality of service (QoS) requirement. Although there has been quite some prior research on the UAV deployment, no work has studied this problem in a 3 dimensional (3D) setting and taken into account the UAV-BS capacity limit and the quality  of  service  (QoS)  requirements of ground users. Therefore, in this paper, we focus on the problem of deploying UAV-BSs to provide satisfactory wireless communication services, with the aim to maximize the total number of covered user equipment (UE) subject to user data rate requirements and UAV-BSs' capacity limit. First, we model the relationship between the air-to-ground (A2G) path loss (PL) and the location of UAV-BSs in both horizontal and vertical dimensions which has not been considered in previous works. Unlike the conventional UAV deployment problem formulation, the 3D deployment problem is decoupled into a 2D horizontal placement and altitude determination connected by path loss requirement and minimization. Then, we propose a novel genetic algorithm (GA) based 2D placement approach in which UAV-BSs are placed to have maximum coverage of the users with consideration of data rate distribution. Finally, numerical and simulation results show that the proposed approach has enabled a better coverage percentage comparing with other schemes.    
\end{abstract}


\IEEEpeerreviewmaketitle

\vspace{-0.3cm}
\section{Introduction}
\IEEEPARstart{U}{nmanned} aerial vehicle (UAV)-assisted \textcolor{black}{communications have} recently gained fast popularity as an effective solution to complement traditional stationary base stations. Unmanned aerial vehicle base stations (UAV-BSs) have the rapid-deployment and reconfiguration advantages compared to terrestrial ones~\cite{Zeng2016}. The roadmap of telecommunication infrastructure provider~\cite{LinUAV} and 3GPP technical reports~\cite{3GPP} have demonstrated promising field trials results of wireless connectivity to the UAVs, and discussed the future ubiquitous mobile broadband coverage both on the ground and in the sky. The \textcolor{black}{fifth generation (5G)} and beyond wireless communications and Internet-of-Things (IoT) application scenarios can be facilitated by the UAV communication or treat it as a critical integral part \cite{Zeng2019a}, \cite{HuoUAV}, \cite{HuoMaritime}. The advantages of high mobility and flexibility of UAVs as part of the high-performance wireless communications network can also potentially serve the broadcasting industry. For example, UAVs are used to conduct object-oriented tracking during aerial filming \cite{DroneFilming} and transmit the broadcasting streams simultaneously. Moreover, a mathematical model was proposed to overcome the problem of sport event filming with connectivity constraints in \cite{EventFilming}. A taxonomy to formulate the concept of multiple-UAV cinematography was proposed to enable the autonomous UAV filming in~\cite{Mademlis2019}.  

Despite the benefits in enabling UAV-BSs in the broadcasting and communications industry, there are significant challenges in terms of UAV system design and deployment strategies. \textcolor{black}{For example, finding suitable UAV-BSs' positions when deploying the UAV-BSs network is particularly difficult in terms of cost-efficiency.} Since the life time of the battery powering one UAV-BS is limited and the number of available UAV-BSs is also constrained, UAV-BSs should be deployed in a method which maximizes the number of covered users in an energy-efficient way. Another critical challenge is that in practical situations, different user equipment (UE) may have different quality of service (QoS) requirements while each UAV-BS has limited data rate capacity. Therefore, the rational distribution of the radio resources needs to be considered.

Research on UAV-BSs development has focused on finding horizontal positioning [10]-[12] and altitude optimization [13]-[15]. In~\cite{Lyu2017} and~\cite{Mozaffari2016}, an identical coverage radius is assumed for all UAV-BSs. The work in~\cite{Lyu2017} proposes an efficient spiral placement algorithm aiming to minimize the required number of UAVs, while~\cite{Mozaffari2016} models the UAV deployment problem based on circle packing theory and study the relationship between the number of deployed UAV-BSs and the coverage duration. In~\cite{Galkin2016}, the authors use a K-Means clustering method to partition the ground users to $k$ subsets and users belonging to the same subset are served by one UAV. All these works have a fixed altitude assumption. The relationship between the altitude of UAV-BSs and the coverage area is studied in~\cite{Al-Hourani2014} and~\cite{Bor-Yaliniz2016}. In~\cite{Al-Hourani2014}, the method of finding the optimal altitude of a single UAV placement for maximizing the coverage is studied based on a channel model with probabilistic path loss (PL). Reference~\cite{Bor-Yaliniz2016} formulates an equivalent problem based on the same channel model as~\cite{Al-Hourani2014} and proposes an efficient solution. Moreover,~\cite{Alzenad2017} studies multiple UAV-BS 3D placements with a given radius taking into account energy efficiency by decoupling the UAV-BS placement in the vertical dimension from the horizontal dimension. \textcolor{black}{In recent years}, artificial intelligence \textcolor{black}{algorithms are exponentially developed and applied in various research fields.}  

In this paper, we investigate multiple 3D UAV-BS deployment with the aim to maximize the number of UAV-served UEs under realistic conditions where each UE has a QoS compliance including a maximum tolerated path loss and a unique data rate requirement and each UAV-BS has a limited sum capacity. \textcolor{black}{The novelty and contributions are summarized as follows:}

\begin{itemize}
\item \textcolor{black}{First, in order to consider a more practical deployment scenario, the QoS compliance of ground users is measured by taking into account the maximum allowed path loss and a unique data rate requirement. It is worth mentioning that the existing problems and results in the literature ignore the QoS requirements, while QoS compliance leads to different coverage radii of UAV-BSs.} 

\item \textcolor{black}{Second, the 3D placement problem is treated as a 2D deployment by placing multiple circles of various sizes in the horizontal dimension and then determining the altitude of UAV-BSs, which simplifies the original problem without losing the accuracy.} 

\item \textcolor{black}{Last, a new genetic algorithm (GA) based UAVs deployment strategy and framework is proposed and proved to provide an effective solution and performance in comparison.}
\end{itemize}

\textcolor{black}{The remainder of this paper is organized as follows. Section II conducts the problem formulation and provides the system model. In Section III, we first analyze the 3D deployment problem and then decouple it into a 2D problem, followed by the determination of the UAVs’ altitudes. In Section IV, the GA algorithm is investigated and analyzed for solving the 2D placement problem. Section V presents the numerical results and discussions. Finally, Section VI concludes the entire paper with the future work. }

\section{System Model}
\label{sec:system_model}
\begin{figure}[t!]
		\centering
		\includegraphics[width=1\columnwidth]{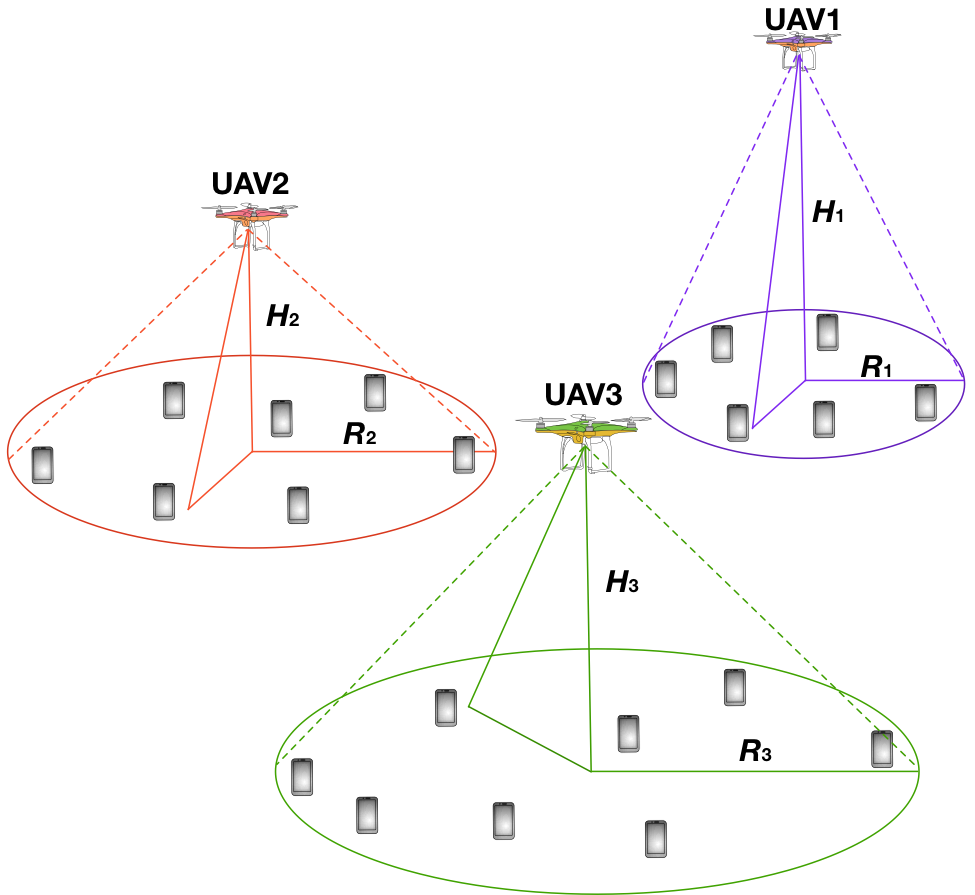}
		\caption{A communication system model of multiple UAV-BSs serving ground users.}
		\label{fig:sys_mol_demo}
		\vspace{-0.4cm}
	\end{figure}

Fig.~\ref{fig:sys_mol_demo} shows a communication network model where many UEs are clustered to be served by multiple UAV-BSs. The objective is to find the optimal locations for UAV-BSs so that the ground users' coverage ratio and the coverage radii can be maximized. Let $\mathcal P$ be the set of all the UEs which are labelled as $i = 1,2,...\left | \mathcal P \right |$. Each UE has a unique data rate requirement ${c_i}$ and all UEs have a maximum tolerated path loss $PL_{max}$ \textcolor{black}{that serves} the purpose to guarantee all the data rate requirements from UEs are feasible, for QoS compliance. $\mathcal Q$ denotes the set of available UAV-BSs labelled as $j = 1,2,...\left | \mathcal Q \right |$ and each UAV-BS has a data rate capacity $C_{j}$. In our system, we assume that no ground base station is available but the locations and data rate requirement of all users are pre-known. \textcolor{black}{Furthermore, in spite of the well-known interference issues in UAV-assisted networks, such as multi-cell co-channel interference \cite{Hou-2019}, \cite{Feng-2019}, this work does not take into account the said interference which can be mitigated by various techniques such as, frequency planning, multi-beam UAV communication scheme \cite{Liu-2018}, mmWave multi-stream multi-beam beamforming~\cite{Huo-VTC-2019}, non-orthogonal multiple access (NOMA) technique~\cite{UAV-System}, cooperative NOMA scheme \cite{Mei-2019}, cooperative interference cancellation strategy \cite{Liu-2018}, \cite{Liu-2019}, and other interference cancellation techniques.}

The A2G channel \textcolor{black}{modeling follows} \cite{Al-Hourani2014} where line-of-sight (LoS) occurs with a certain probability, \textcolor{black}{which falls into our application scenarios}. The probability of a LoS and non line-of-sight (NLoS) channel between UAV $j$ at the horizontal position $m_{j}=(x_{j},y_{j})$ and user $i$ at the horizontal location $u_{i}=(\tilde{x}_{i},\tilde{y}_{i})$ are formulated as~\cite{Al-Hourani2014}
\begin{equation}
\begin{aligned}
\label{eqn:prob}
P_{LoS}=&\frac{1}{1+a\exp (-b(\frac{180}{\pi }\tan^{-1}(\frac{H_{j}}{r_{ij}})-a))},\\
P_{NLoS}=&1-P_{LoS},
\end{aligned}
\end{equation}
where $H_{j}$ is the altitude of UAV-BS $j$; $a$ and $b$ are environment dependent variables; $r_{ij} = \sqrt{(x_{j}-\tilde{x}_{i})^{2}+( y_{j}-\tilde{y}_{i} )^{2}}$ is the horizontal euclidean distance between the $i^{th}$ user and $j^{th}$ UAV. Then the path loss for LoS and NLoS can be written as 
\begin{equation}
\begin{aligned}
\label{eqn:loss}
PL_{LoS}=20\log (\frac{4\pi f_{c}d_{ij}}{c})+\eta _{LoS},\\PL_{NLoS}=20\log (\frac{4\pi f_{c}d_{ij}}{c})+\eta _{NLoS}
\end{aligned}
\end{equation}
where $f_{c}$ is the carrier frequency, $c$ is the speed of light and $d_{ij}$ denotes the distance between between the UE and UAV-BS given by $d_{ij}=\sqrt{H_{j}^{2}+r_{ij}^{2}}$. Moreover, $\eta _{LoS}$ and $\eta _{NLoS}$ are the environment dependent average additional path loss for LoS and NLoS condition respectively. According to~\eqref{eqn:prob},~\eqref{eqn:loss}, the path loss (PL) can written as:
\begin{equation}
\begin{aligned}
\label{eqn:PL}
PL=&PL_{LoS}\times P_{LoS}+PL_{NLoS}\times P_{NLoS}\\
=&\frac{A}{1+a\exp (-b(\frac{180}{\pi }( \frac{H_{j}}{r_{ij}})-a))}+20\log \frac{r_{ij}}{\cos (\frac{H_{j}}{r_{ij}})}+B
\end{aligned}
\end{equation}
where $A=\eta _{LoS}-\eta _{NLoS}$ and $B = 20\log (\frac{4\pi f_{c}}{c})+\eta _{NLoS}$.

In order to show the effect of different $PL_{max}$ on the radius-altitude curve, we have plotted this relation \eqref{eqn:PL} in Fig.~\ref{fig:atg} where the coverage radius is a function of both,
the altitude $H$ and the $PL_{max}$, by keeping the \textcolor{black}{urban environment statistical parameter set as (9.61, 0.43, 0.1, 20).}

\section{UAV-BS 3D deployment problem}
\label{sec:problem}
\begin{figure}[t!]
		\centering
		\includegraphics[width=3.8in,height=3.2in]{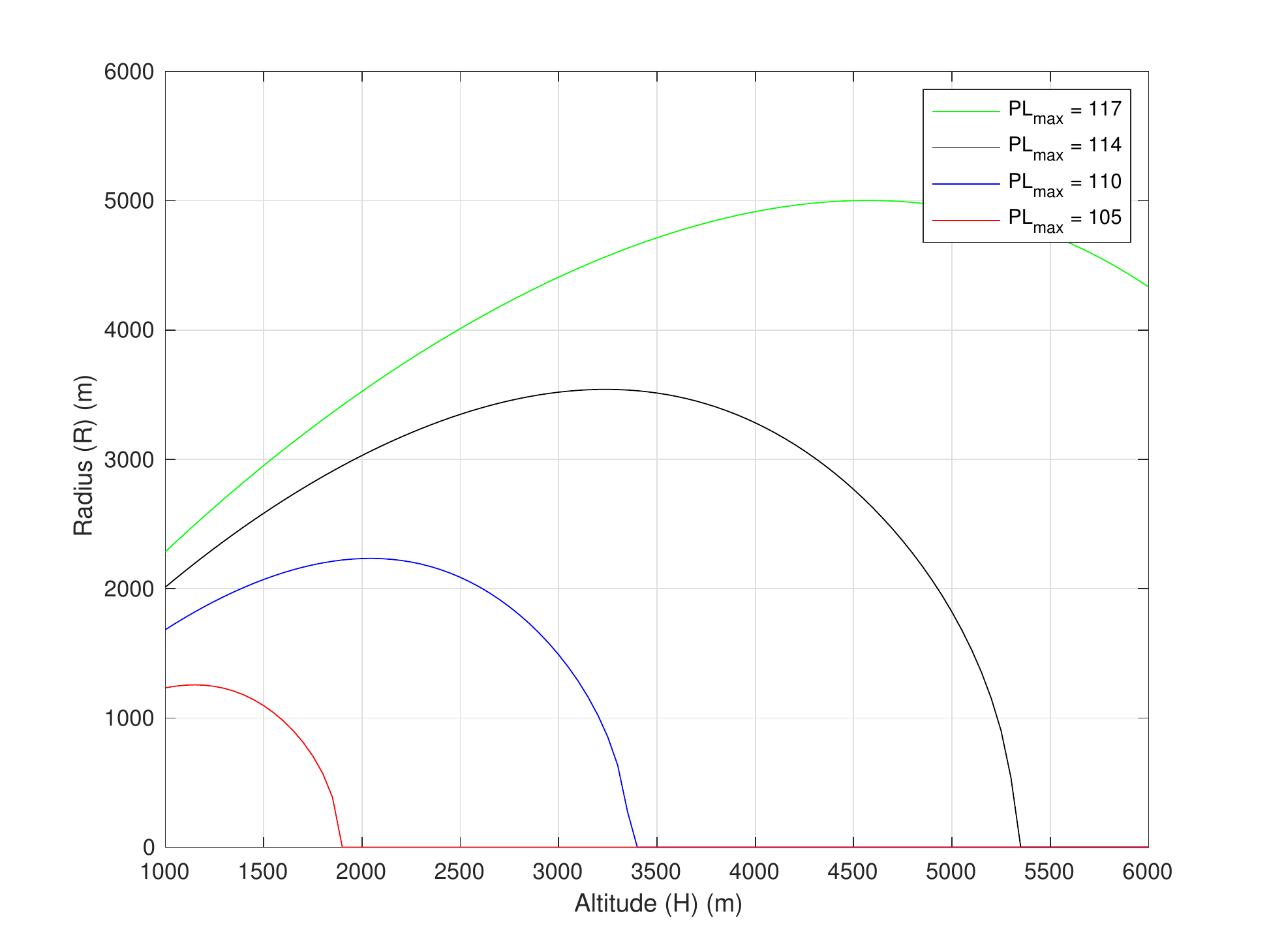}
		\caption{Radius vs. altitude curve for different maximum path loss.}
		\label{fig:atg}
	\end{figure}

For the problem at hand, the 3D deployment of UAV-BSs can be decomposed into the 2D horizontal locations optimization and altitude determination. This is because the UAV altitude only impacts the cell radius and path loss experienced in the cell, while the horizontal location and a radius determine which UEs are covered by the UAV. As clearly seen in Fig.~\ref{fig:atg}, for a given $PL_{max}$, there is a maximum radius $R_{max}$ and a corresponding altitude $H_{max}$. If the altitude is smaller or larger than $H_{max}$, while maintaining the same radius, the path loss on the cell edge will be larger than the given $PL_{max}$. Since the cell radius affects the total number of the covered UEs, we want the cell radius to be maximized in order to potentially cover more users. Hence the 3D deployment solution takes the procedure as follows. First, a maximum cell radius upper bound $R_{max}$ that guarantees the desired $PL_{max}$ requirement is derived. Second, the 2D placements of $\left | Q \right |$ UAVs and their respective coverage radii bounded by $R_{max}$ that maximize the total number of UEs supported while satisfying the individual data rate requirements and the UAV capacity constraint are formulated and solved. Finally, given the actual coverage radius of each UAV obtained from the second step, the altitude that leads to the achieved minimum cell edge path loss is determined.  

\subsection{2D UAV-BS Deployment Problem}
\label{sec:a}
Since we model the 2D deployment problem via placing multiple circles of different sizes, unlike authors in~\cite{Al-Turjman2019} who investigate a problem of solving for the least number of UAVs to cover users in a region, this problem is equivalent to finding the appropriate location and radius for each UAV-BS to cover as many UEs as possible while simultaneously satisfying the data rate requirements and the UAV capacity constraint.

A binary variable $\gamma _{ij}\in \left \{ 0,1 \right \}$ is used to indicate whether or not the user $i$ is covered by UAV-BS $j$, 1 for service and 0 for no coverage. The necessary condition for user $i$ to be covered by UAV-BS $j$ is that the horizontal Euclidean distance between them is less than the coverage radius of UAV-BS $j$, $R_{j}$, which can be written as $\left \| m_{j}-\gamma _{ij}u_{i} \right \|\leq R_{j}$. Following~\cite{Mozaffari2016}, the constraint equation can be rewritten as \textcolor{black}{ $\left \| m_{j}-\gamma _{ij}u_{i} \right \|\leq R_{j}+M(1-\gamma _{ij})$,} where $M$ is a large constant which is larger than the largest horizontal distance between a user and a UAV so the constraint holds in any condition. \textcolor{black}{As defined earlier, $m_{j}=(x_{j},y_{j})$ and it stands for the horizontal position of UAV $j$ while $u_{i}=(\tilde{x}_{i},\tilde{y}_{i})$ represents the horizontal location of user $i$.}

If a user is within the serving area of a UAV-BS, the UAV-BS can allocate certain data channels to the user which has a unique data rate requirement $c_{i}$. For simplicity, we assume that for any UE, the allocated data rate equals what it requires. Then the data rate allocation problem can be expressed as \textcolor{black}{$\sum_{i=1}^{\left | \mathcal P \right |}c_{i}\gamma_{ij}\leq C_{j},j\in \left \{ 1,2,...\left | \mathcal Q \right | \right \}$, \text{where}~$C_{j}$ is the data capacity of UAV $j$.}

At this stage, the UAV deployment problem becomes a rucksack-like problem in combinatorial optimization, which is a NP-hard problem. It can be expressed as
\begin{equation}
	\begin{aligned}
	&\underset{ R_{j},m_{j}}{\text{maximize }}\sum_{j=1}^{\left | \mathcal Q \right |}\sum_{i=1}^{\left | \mathcal P \right |}\gamma _{ij},\\
	\text{s.t. }
	C1:&\left \| m_{j}-\gamma _{ij}u_{i} \right \|\leq R_{j}+M(1-\gamma _{ij}),\\
	&i\in \left \{ 1,2,...\left | \mathcal P \right | \right \},j\in \left \{ 1,2,...\left | \mathcal Q \right | \right \},\gamma _{ij}\in \left \{ 0,1 \right \}\\
	C2:&\sum_{i=1}^{\left | \mathcal P \right |}c_{i}\gamma_{ij}\leq C_{j},j\in \left \{ 1,2,...\left | \mathcal Q \right | \right \}\\
	C3:&\sum_{j=1}^{\left | \mathcal Q \right |}\gamma _{ij}\leq 1,i\in \left \{ 1,2,..\left | \mathcal P \right | \right \}\\
	C4:&R_{j}\leq R_{max},j\in \left \{ 1,2,...\left | \mathcal Q \right | \right \}.
	\end{aligned}\label{eqn:optimize}
\end{equation}

Our objective is to maximize the number of served users. First, $C1$ in~\eqref{eqn:optimize}, guarantees that a UE can be served by a UAV-BS, when the horizontal distance between the UE and the UAV-BS is less than UAV-BS's coverage radius. Then $C2$ regulates that the total data rate of all covered users served by one UAV-BS cannot exceed the data rate capacity of the UAV-BS. Furthermore, $C3$ ensures each user should be served by at most one UAV-BS. Last, Fig.~\ref{fig:atg} shows that the function of coverage radius respective to altitude for a given $PL_{max}$ is a concave function so there exists a maximum radius $R_{max}$ that any coverage radii $R > R_{max}$ does not have a feasible solution. Thus, $C4$ ensures that the radii of UAV-BSs are no larger than $R_{max}$. A genetic algorithm to solve this optimization problem will be presented in the next section.

\subsection{The Determination of UAV-BS Altitude}
After Subsection~\ref{sec:a}, the horizontal locations and coverage radii of UAV-BSs have been determined and all the coverage radii are less than $R_{max}$. Therefore, for each UAV-BS, the range of altitude which results in  the $PL$ value less than $PL_{max}$ can be obtained from Fig.~\ref{fig:atg}. The objective for this step is to find the optimal altitude for each UAV-BS which requires least transmit energy, ie., the minimum path loss, to provide service for the coverage range derived in step 1.

\begin{figure}[t!]
		\centering
		\includegraphics[width=3.9in,height=3.2in]{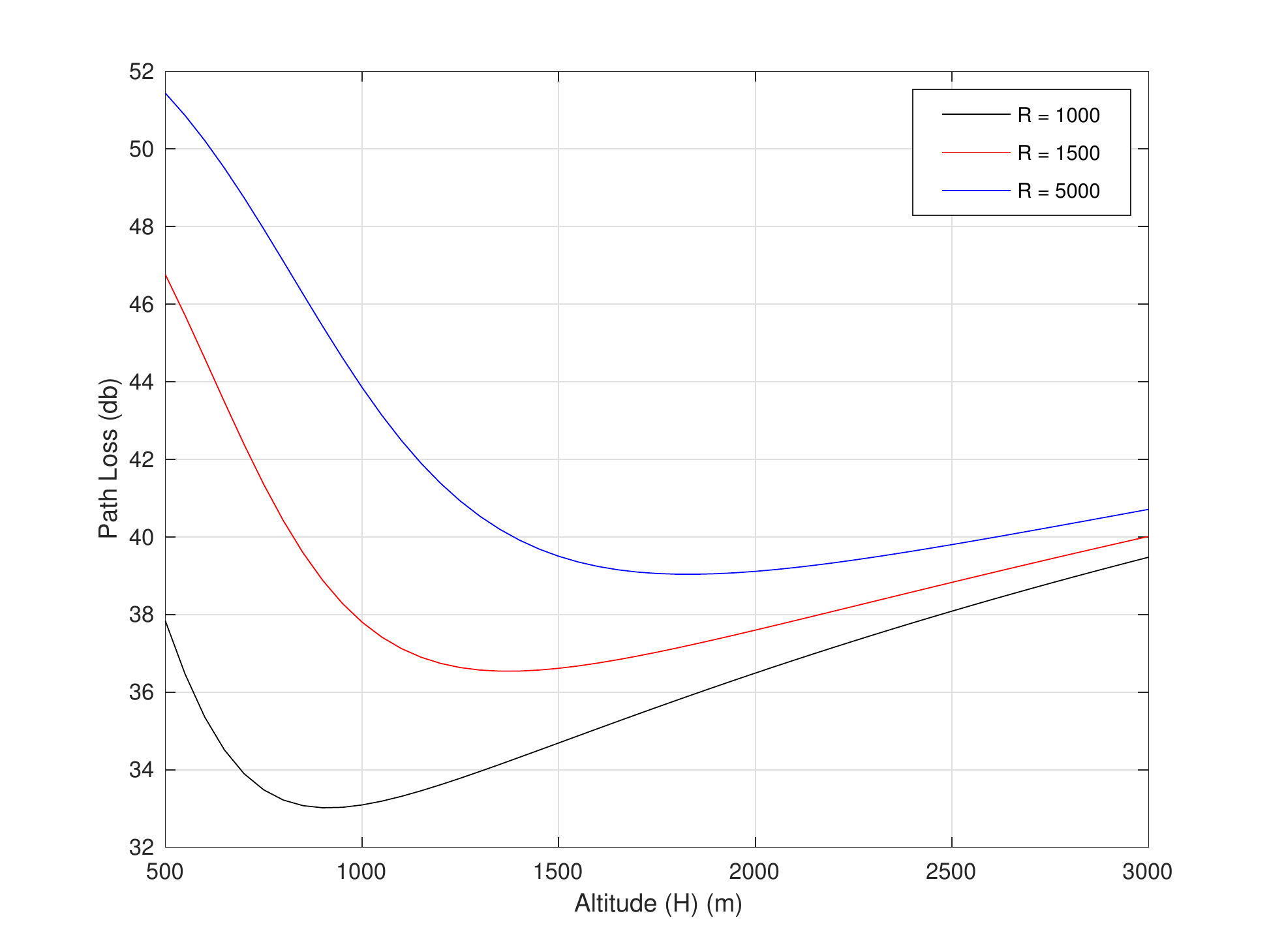}
		\caption{Path loss vs. altitude for given radii in an urban environment.}
		\label{fig:altitude}
	\end{figure}

As observed from~\eqref{eqn:PL}, the path loss between a UAV-BS and UE is a function of the horizontal distance $r$ and the altitude $H$, that is, $PL=f(r, H)$. Also, from Fig.~\ref{fig:atg}, for a given $PL_{max}$, defining the elevation angle $\theta = \frac{H}{R}$, there exists an elevation angle $\theta_{max}$ that maximizes the radius $R$ by solving \textcolor{black}{ $\frac{\partial R}{\partial H}=0$}. As derived in~\cite{Al-Hourani2014}, $\theta_{max}$ satisfies the following equation:
\begin{equation}
\begin{aligned}
\label{eqn:opt_angle}
\frac{\pi }{9\ln (10)}\tan (\theta _{max})+\frac{abA\exp (-b(\frac{180}{\pi }\theta _{max}-a))}{(a\exp (-b(\frac{180}{\pi }\theta _{max}-a))+1)^{2}}=0
\end{aligned}
\end{equation}
where $\theta_{max}$ is environment dependent so it is a constant in a given environment. It has been proven by~\cite{Alzenad2017} that this elevation angle provides the minimum $PL$ of the users in the boundary which is equivalent to the $PL$ of all the UEs within the covered range are minimized so the required transmit power of the UAV-BS is minimized. Therefore, once the actual coverage radius $R$ of each UAV-BS is obtained in Subsection III-A, the UAV-BS altitude $H_{opt}$ is given by $H_{opt}=R\tan (\theta _{max})$. Fig.~\ref{fig:altitude} shows the relationship between $PL$ and altitude for given radii. It can be observed that as long as the radius is fixed, a minimum value of $PL$ always exists.


\section{Genetic Algorithm based 2D Placement}
\label{sec:algorithm}

\textcolor{black}{ In order to solve complex optimization problems, there have been a wide range of applications of swarm intelligence algorithms \cite{Slowik2018}, \cite{Govindan}, \cite{Zhao_2019},\cite{Brezocnik},
\cite{Anandakumar} and evolutionary-based metaheuristics \cite{Dulebenets}, \cite{Dulebenets_2019}, \cite{Vahdani}. For example, \cite{Slowik2018} presented a detailed analysis on evolutionary algorithm based real-life applications.} In this section, we present a GA based UAV-BS deployment strategy to provide wireless services for a group of UEs. The objective is to solve the optimization problem \eqref{eqn:optimize}. \textcolor{black}{ The genetic algorithm is an efficient solution to the complex optimization problems with multiple variables, widely appearing in real-life optimization problems of a variety of fields. For example, \cite{Yang2020} introduced a method based on GA and deep learning to predict financial behaviours. Moreover, the genetic algorithm was indeed applied in the cellular communications related research field such as facilitating terrestrial base station placement and showed excellent efficiency \cite{GA-BS-2019}. }

GA works on a population which consists of some candidate solutions and the population size is the total number of solutions~\cite{Goldberg:1989:GAS:534133}. Each solution is considered to be a chromosome and each chromosome has a set of genes where each gene is represented by the features of the solutions. Then, each individual chromosome has a fitness value which is computed based on the fitness function representing the quality of the chromosome. Moreover, a selection method called roulette  wheel  method where the chromosome with higher fitness value has a higher chance to survive the population.

\begin{figure}[t!]
		\centering
		\includegraphics[width=0.5\columnwidth]{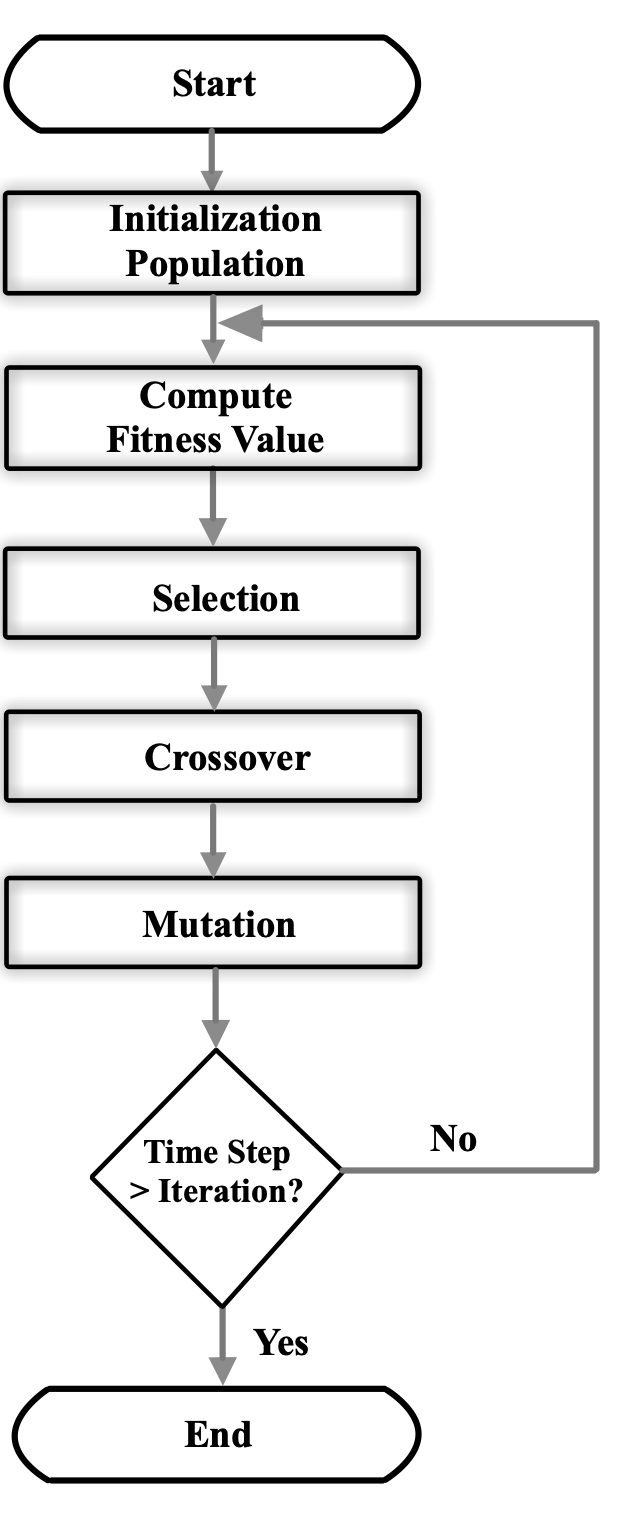}
		\caption{Genetic algorithm workflow.}
		\label{fig:GA}
	\end{figure}

\textcolor{black}{However, the selection process can only assure in each generation, a better solution has a higher chance to enter the next generation.} In order to ensure the diversity of the solution to avoid falling into local optimal solutions, crossover and selection are applied after the selection process. In the crossover procedure, two chromosome are selected in a probability of crossover rate to exchange information so new chromosomes are generated. Also, in the mutation procedure, each chromosome has a probability of mutation rate to \textcolor{black}{replace} a set of genes with new random values. \textcolor{black}{The basic GA process has various variables including population size, crossover rate and mutation rate. The population size determines the size of candidate solutions. The value of crossover rate and mutation rate represent the diversity of the candidate solutions throughout the iteration.} The whole process repeats until the time step reaches an iteration limit. Fig. \ref{fig:GA} illustrates the whole process of GA.

As illustrated in Algorithm \ref{alg:proposed_algorithm}, the horizontal location, and the coverage radius of each UAV-BS are treated as a gene in the GA model. Therefore, for UAV-BS $j$, the combination ($x_{j}$, $y_{j}$, $R_{j}$) is a gene. Placing genes for all the available UAV-BSs together, i.e., $\left \{ x_{j}, y_{j}, R_{j} \right \}_{j\in \mathcal Q}$ makes a chromosome. The required inputs include $K$, $D$, $\mathcal P$, $\mathcal Q$, $R_{max}$, $\left \{ c_{i} \right \}_{i\in \mathcal P}$, $\left \{ u_{i} \right \}_{i\in \mathcal P}$, $\theta _{opt}$, $p_{m}$, $p_{c}$ where $K$ is the number of iterations for finding the optimal result, $D$, $p_{m}$ and $p_{c}$ are the population size, mutation rate and crossover rate for GA respectively. The outputs are the horizontal locations, altitudes and coverage radii, denoted by $O_{j}, j = 1,2,...\left | \mathcal Q \right |$, of all the UAV-BSs. 

First, $\left | \mathcal Q \right |$  empty lists are created and each of them is to store the covered UEs of the corresponding UAV-BSs. Also, two arrays $r$, $\hat{r}$ are created, respectively, to store the number of covered UEs in each UAV-BS and the total number of covered UEs of all UAV-BSs known as the fitness score. In step 3, the first population $\nu _{1}$ is generated by creating $D$ chromosomes where the horizontal locations of all UAV-BSs are initialized by assigning each of them with the equidistant point of 3 random UEs' locations, and the coverage radius \textcolor{black}{is} initialized by generating a random numbers in the range from 1 to $R_{max}$. 

Then, $K$ iterations are executed to find the 2D deployment result from Step 4 to Step 20. In Step 5 and Step 6, if the horizontal distance between a UE and a UAV-BS is less than the coverage radius, the UE can be served by the UAV-BS. Also, if a UE is within the coverage range of more than one UAV-BS, it is assigned to the closest one. In the for loop from Step 7 to Step 16, calculate the sum data rate $\sum_{\hat{p}\in \mathcal O_{j}}^{} c_{\hat{p}}$ of all covered UEs for each UAV-BS. If the sum data rate is smaller than the data capacity $C_{j}$, the number of covered UEs $\left | \mathcal O_{j} \right |$ is stored to array $r$. Otherwise, a negative number is stored to array $\hat{r}$ and the algorithm breaks out the loop and goes back to Step 5, which means the fitness of this chromosome is negative. In Step 15, the fitness function of the chromosome is the total number of covered UEs and it is saved into array $\hat{r}$. 

In Step 17, the roulette wheel method is applied to update the current population $\nu _{\hat{k}}$. A random chromosome is selected within the current population to be the competitor. Comparing the fitness score of all the chromosomes with the competitor, the chromosomes with less fitness scores are replaced by the competitor. Afterward, in the crossover procedure, $p_{c}$ of chromosomes are randomly selected and paired. Each pair is considered to be the parent chromosomes. In each parent chromosomes, the first half genes of one chromosome and second half genes of the other chromosome are exchanged to produce children chromosomes. In Step 19, all the chromosomes have a probability of $p_{m}$ to perform mutation process in which one gene of the mutated chromosome is selected to be replaced by random horizontal location and coverage radius.

Finally, in Step 21 and Step 22, we can obtain the result of horizontal locations and coverage radii of UAV-BSs via choosing the chromosome with the maximum fitness score. Finally, the optimal altitudes are obtained by $H_{opt}=R\tan (\theta _{max})$.

\textcolor{black}{Since we model our problem to be a rucksack problem and put additional constraints on it, the computational complexity is proportional to the search space. In our proposed algorithm, Step 4 to Step 16 have complexity $\mathcal{O}(KLD)$ where $K$ is the number of UAV-BSs, $L$ is the iteration time, $\mathcal D$ is the number of chromosomes. Step 17 to 19 perform the mathematical operation and cost $\mathcal{O}(1)$, and similarly, step 21 and 22 also cost $\mathcal{O}(1)$. Therefore, the time complexity of the GA UAV-BS deployment method is $\mathcal{O}(KLD)$.}

\begin{algorithm}[t!]
{\footnotesize\caption{\footnotesize Genetic Algorithm Based 2D UAV-BS Placement Method}
\label{alg:proposed_algorithm}
    \hspace*{\algorithmicindent} \textbf{Input: }$K$, $D$, $\mathcal P$, $\mathcal Q$, $\left \{ c_{i} \right \}_{i\in \mathcal P}$, $\left \{ u_{i} \right \}_{i\in \mathcal P}$, $\theta _{max}$, $p_{m}$, $p_{c}$  \\
    \hspace*{\algorithmicindent} \textbf{Output:} $\left \{ m_{j} \right \}_{j\in \mathcal Q}$,  $\left \{ H_{j} \right \}_{j\in \mathcal Q}$,  $\left \{ R_{j} \right \}_{j\in \mathcal Q}$
	\begin{algorithmic}[1]
		\STATE Create $\left | \mathcal Q \right |$ empty list $\left \{ O_{j} \right \}_{j\in \mathcal Q}$. 
		\STATE Create two arrays $r$, $\hat{r}$ with size $\left | \mathcal Q \right |$ and $D$ respectively.
		\STATE \textbf{Initialize Population:} Initialize first population $\nu _{1}$ by creating $D$ sets of chromosomes.
		\FOR {$\hat{k} =1$; $\hat{k}\leq K$; $\hat{k}++$ }
		\FOR {$\hat{i} =1$; $\hat{i}\leq D$; $\hat{i}++$ }
		\STATE Perform UE assignments according to their distances to UAV-BSs and store the results in $\left \{ O_{j} \right \}_{j\in \mathcal Q}$.
		\FOR{$j =1$; $j\leq \left | \mathcal Q \right |$; $j++$}
        \IF{$\sum_{\hat{p}\in O_{j}}^{} c_{\hat{p}}\leq C_{j}$} 
        \STATE $r\left [ j \right ]\leftarrow \left | O_{j} \right |$
        \ELSE 
        \STATE $\hat{r}[\hat{i}]\leftarrow -100$ 
        \STATE Continue and go back to step 5
        \ENDIF 
		\ENDFOR
		\STATE \textbf{Fitness Function:} $\hat{r}[\hat{i}] = sum(r)$
		\ENDFOR
		\STATE \textbf{Selection:} update $\nu _{\hat{k}}$ using roulette wheel method to select chromosomes
		\STATE \textbf{Crossover:} Based on $p_{c}$, update $\nu _{\hat{k}}$ by exchanging information of parent chromosomes to produce children chromosomes 
		\STATE \textbf{Mutation:} Based on $p_{m}$, gene is selected randomly to replace new random values 
		\ENDFOR
		\STATE Find the chromosome with maximum value in $\hat{r}$ and obtain $\left \{ m_{j} \right \}_{j\in \mathcal Q}$ and $\left \{ R_{j} \right \}_{j\in \mathcal Q}$ from the chromosome.
		\STATE Obtain $\left \{ H_{j} \right \}_{j\in \mathcal Q}$ by solving $H=R\tan (\theta _{max})$
		\RETURN $\left \{ H_{j} \right \}_{j\in \mathcal Q}$, $\left \{ R_{j} \right \}_{j\in \mathcal Q}$, $\left \{ m_{j} \right \}_{j\in \mathcal Q}$
\end{algorithmic}}
\end{algorithm}
\section{numerical results}
\label{sec:experiment}
	\begin{figure}[t!]
		\centering
		\includegraphics[width=3.8in,height=3.5in]{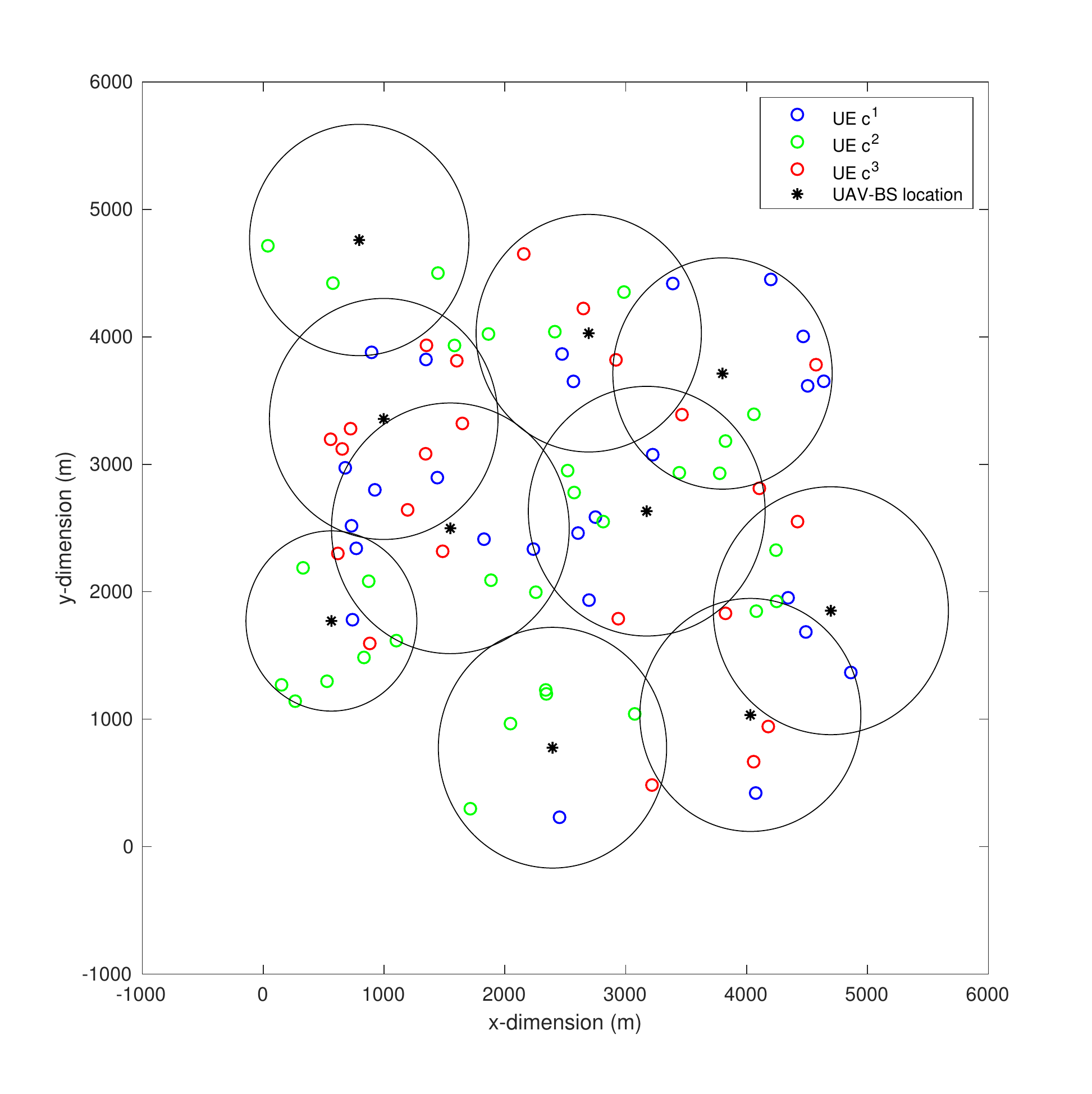}
		\caption{The 100\% coverage ratio result of GA deployment with 80 UEs in a 5000 m $\times $ 5000 m square region with different data rate requirements.}
		\label{fig:deployment}
	\end{figure}

	\begin{figure}[t!]
		\centering
		\includegraphics[width=3.8in,height=3.2in]{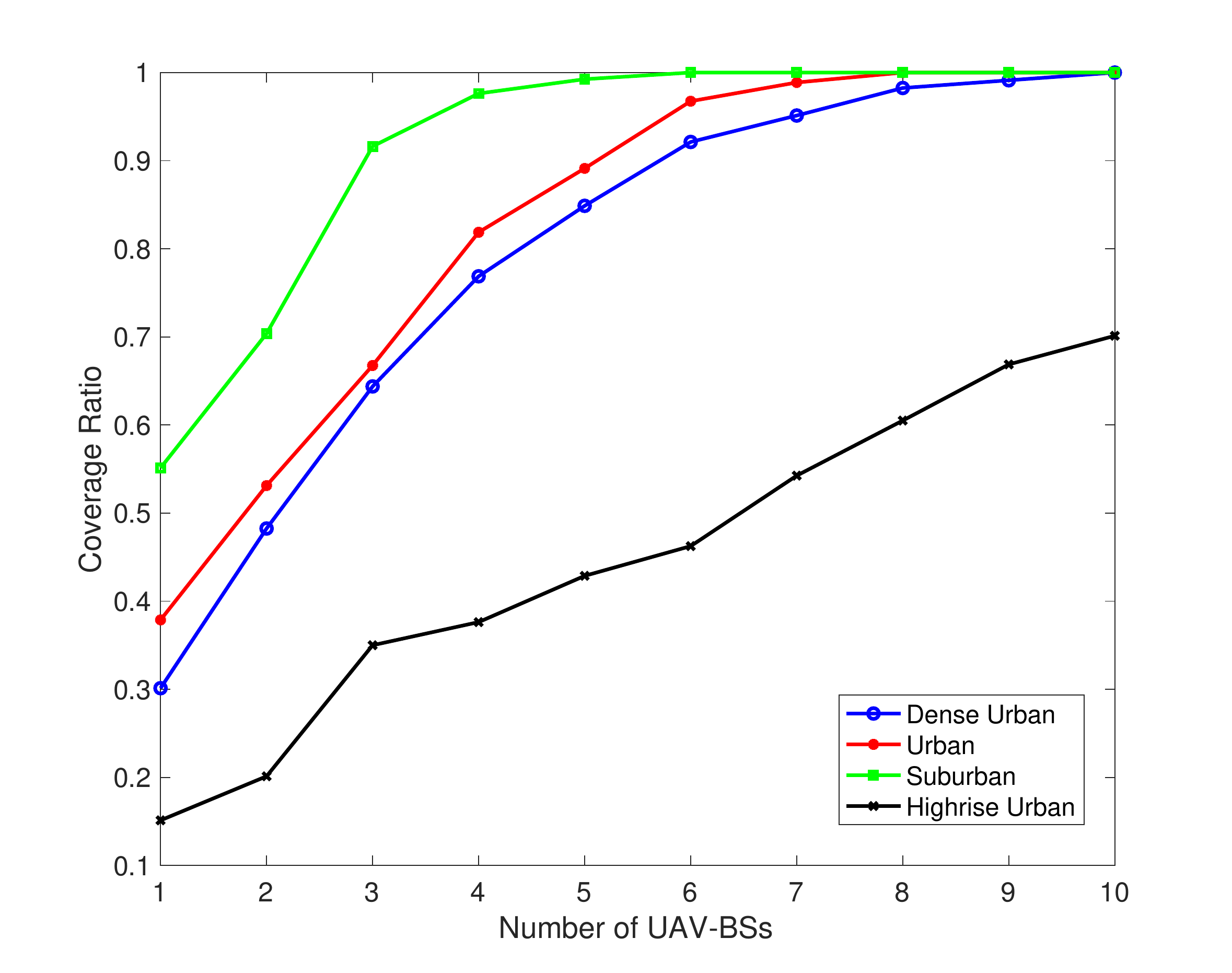}
		\caption{The coverage ratio versus the number of UAV-BSs in four environments.}
		\label{fig:sys_mol_data}
		\vspace{-0.4cm}
	\end{figure}

In our simulations, we consider the UEs are \textcolor{black}{uniformly distributed} in a 5000 m $\times$ 5000 m area. Referring to~\cite{Al-Hourani2014}, the environment parameters are set up as followed: $f_{c}$ = 2 GHz, $PL_{max}$ = 110 dB, ($a$, $b$, $\eta _{LoS}$, $\eta _{NLoS}$) is configured to be (4.88, 0.43, 0.1, 21), (9.61, 0.43, 0.1, 20), (12.08, 0.11, 1.6, 23), (27.23, 0.08, 2.3, 34) corresponding to suburban, urban, dense urban and high-rise urban environments, respectively. Also, we assume there are three different data rate requirements of all UEs, $c^{1}=5\times 10^{6}$ bps, $c^{2}=2\times 10^{6}$ bps and $c^{3}=1\times 10^{6}$ bps, and each UE has one of these three data rate requirements. Moreover, all the UAV-BSs have the same data rate capacity $C=1\times 10^{8}$ bps. Fig. \ref{fig:deployment} illustrates the UE distribution and the GA deployment result  with 100\% coverage percentage.

\begin{figure}[t!]
		\centering
		\includegraphics[width=3.8in,height=3.2in]{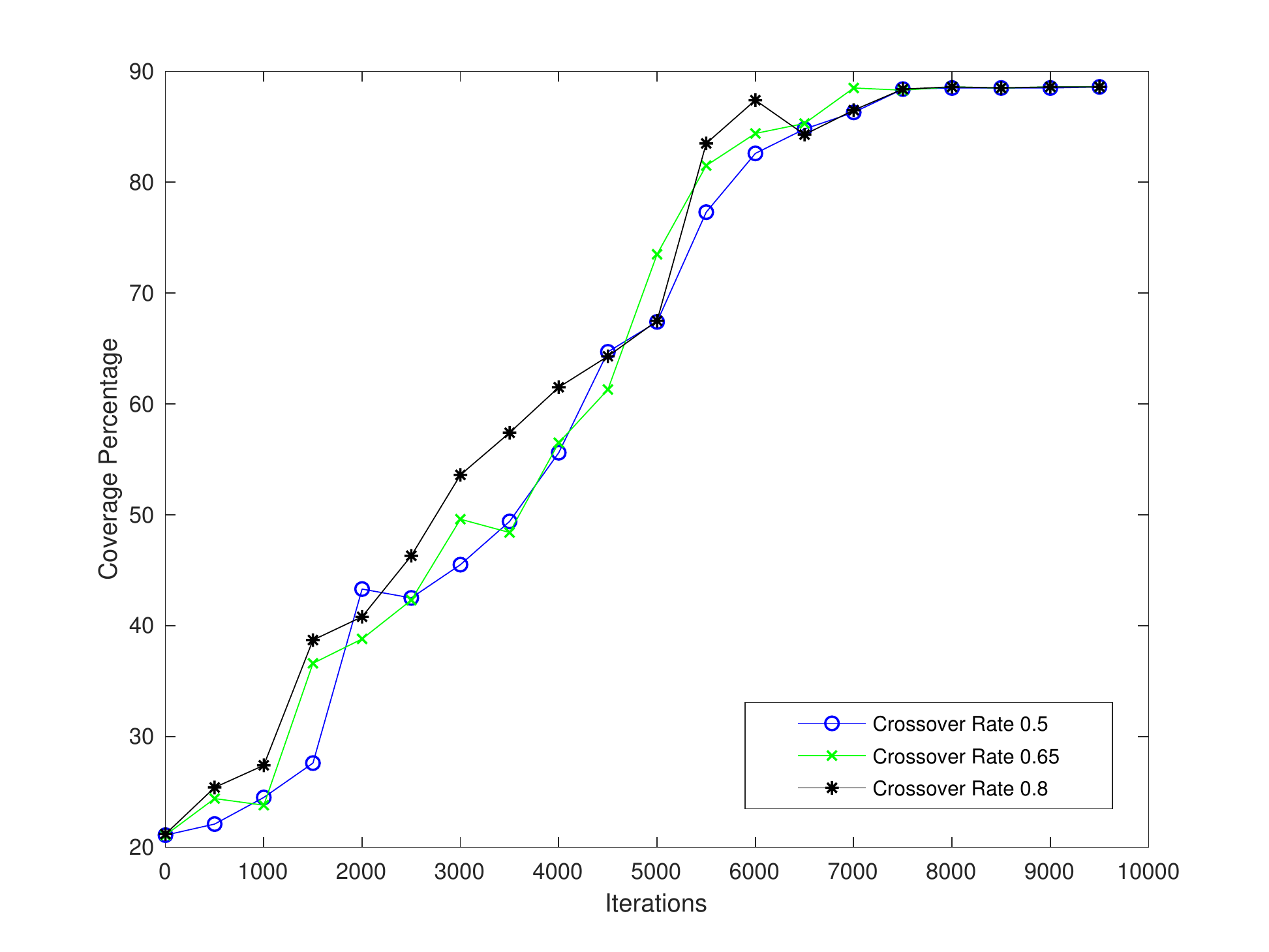}
	    \caption{	\textcolor{black}{GA iteration with different crossover rates.}}
		\label{fig:crossover}
		\vspace{-0.4cm}
	\end{figure}

\begin{figure}[t!]
		\centering
		\includegraphics[width=3.8in,height=3.2in]{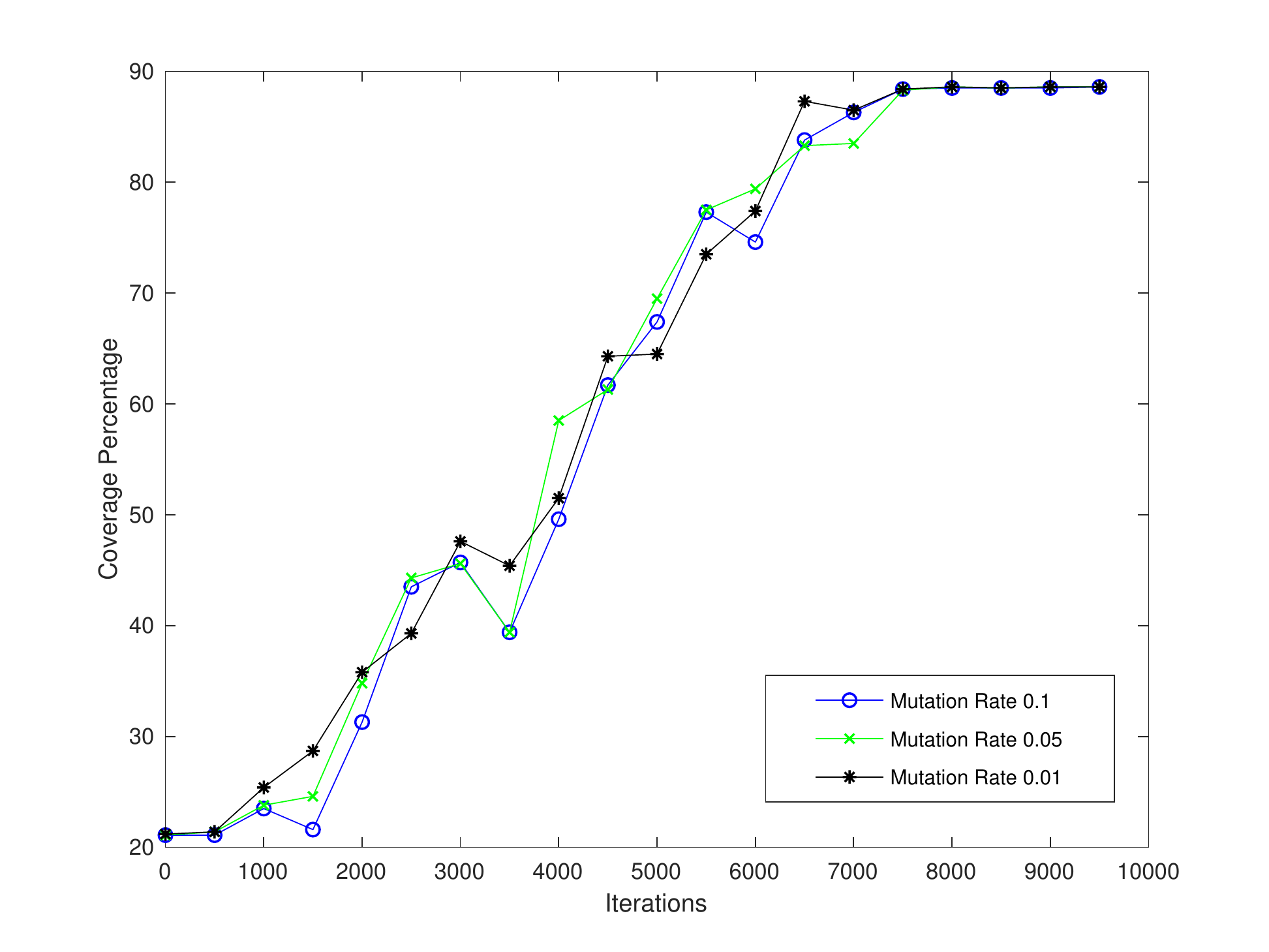}
		\caption{\textcolor{black}{GA iteration with different mutation rates.}}
		\label{fig:mutation}
		\vspace{-0.4cm}
	\end{figure}

\textcolor{black}{In our optimization problem, there are four variables which we need to set up, which are population size, iteration time, mutation rate and crossover rate. According to\cite{Herrera2001}, the range of crossover rate and mutation rate are within [0.5, 0.8] and [0.01, 0.05], respectively. In our problem, in order to analyze how those two parameters affect the algorithm efficiency, in the simulation we fix the iteration size and the population size to be 10000 and 100 respectively, and deploy 10 UAV-BSs to cover 200 UEs. As a result, Fig. \ref{fig:crossover} and Fig. \ref{fig:mutation} show that in our optimization problem those two parameters hardly have impact on the efficiency of the convergence. Furthermore, the parameters ($p_{m}$, $p_{c}$) are configured to be (0.01, 0.8). The time complexity of GA is related to the multiplication of population size and iteration size, as mentioned in Section IV. In other words, these two parameters have a significant impact on the algorithm efficiency where a smaller the multiplication results in a higher GA efficiency. Thus, we conduct an analysis of the relation between the population size and the minimum required iteration number. The minimum required iteration number is defined to be the number of iterations taken to get the fitness value converged to a certain number for 15\% of the entire iteration number. Moreover, Table~\ref{table:1} shows that setting population size to be 100 paired with a iteration number of 17000 can make the GA demonstrate good performance in terms of efficiency. Therefore, the GA parameters set ($K$, $D$, $p_{m}$, $p_{c}$) is configured to be (17000, 100, 0.01, 0.8).} 

\begin{table}[h!]
\vspace{0cm}
\centering
\caption{\textcolor{black}{Multiplication of the population size and the iteration number.}}
\begin{tabular}{||c c c ||} 
 \hline
 Population Size & Minimum Required Iteration & Multiplication \\ [-0.5ex] 
 \hline\hline
 50 & 35380 &  1769000 \\ 
 75 & 23031 & 1727325 \\
 100 & 16875 & 1687500 \\
 150 & 15984 & 2397600 \\
 200 & 16036 & 3207200 \\
 300 & 15944 & 4783200 \\
 500 &16015 & 8007500 \\
 [0ex] 
 \hline
\end{tabular}
\label{table:1}
\end{table}

Fig. \ref{fig:sys_mol_data} shows the average coverage ratios of 80 UEs by \textcolor{black}{10} available UAV-BSs with 15 realizations in four different environments when increasing the number of UAV-BSs. As seen from Fig. \ref{fig:sys_mol_data}, the coverage ratio varies significantly in four deployment scenarios, particularly with high-rise urban one much more challenging than others. 

\begin{figure}[t!]
		\centering
		\includegraphics[width=3.8in,height=3.3in]{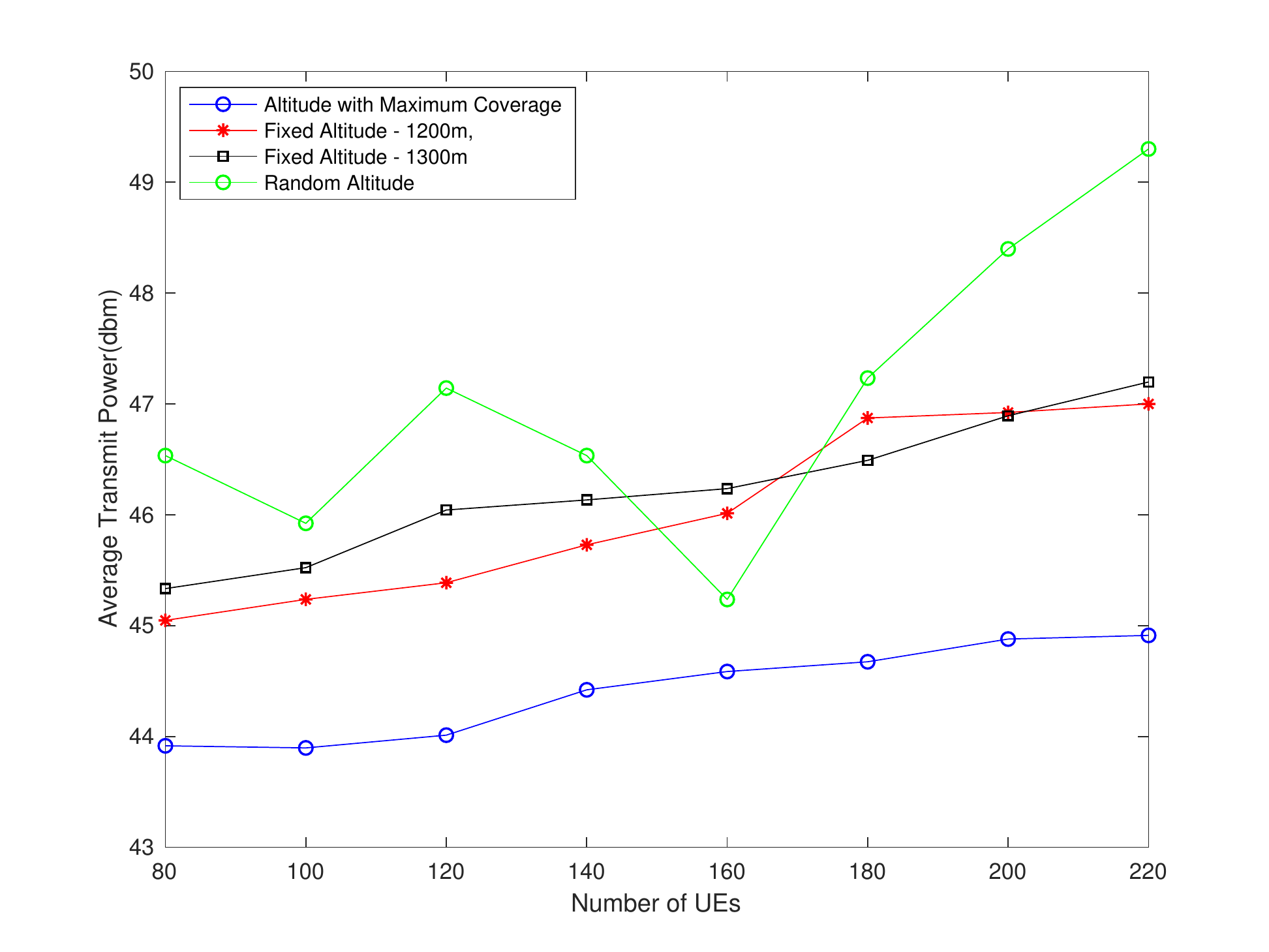}
		\caption{\textcolor{black}{The UAV's average transmit power comparison of altitude with maximum coverage, fixed altitude and random altitude in urban environments.}}
		\label{fig:power}
	\end{figure}

By applying Shannon Capacity Theorem, the required $SNR$ of each UAV-BS can be calculated through $C = B\log _{2}(1 + \frac{P_{r}}{P_{n}})$, where $B$ is the bandwidth of the channel, $P_{r}$ and $P_{n}$ denote the required received power and average noise power, respectively. In our model, we assume that $B=1\times 10^{7}$ Hz, $P_{r} = -74~$\textcolor{black}{dBm} and $P_{n} = -100$ dBm. Thus, we can obtain the minimum required power for each UAV-BS by $P_{t} = P_{r} + PL(R_{j},H_{j})$. \textcolor{black}{Fig. \ref{fig:power} further depicts the average minimum required transmit power of all UAV-BSs when increasing the number of UEs, in the urban environment, with \textcolor{black}{15} available UAV-BSs, and 4 different approaches that determine altitudes. In the fixed altitude approach, all the UAV-BSs are
deployed in the same altitude. In the random altitude approach, each UAV-BS is deployed at an altitude that is uniformly drawn from a feasible range.}  The altitudes from both fixed altitude and random altitudes are selected from the range where $PL_{max}$ requirement is met. As we can see, if the UAV-BSs are deployed in the altitude in the way we proposed, less average transmit power is required to provide wireless service.

\begin{table}[h]
\vspace{0cm}
\centering
\caption{\textcolor{black}{Coverage ratio comparison in urban environments.}}
\begin{tabular}{||c c c c||} 
 \hline
 \text{Methods--}{$\left | \mathcal P \right | $}  & 80 & 200 & 450 \\ [-0.5ex] 
 \hline\hline
 GA Deployment Method & 99.2\% & 88.6\% & 75.3\% \\ 
 K-Means & 98.6\% & 82.3\% & 69.4\% \\
 Branch and Cut  & 95.6\% & 83.1\% & 69.4\% \\
 Greedy Search  & 92.6\% & 79.1\% & 71.4\% \\
 Random & 85.6\% & 72.1\% & 59.4\% \\
 [0ex] 
 \hline
\end{tabular}
\label{table:2}
\end{table}

For further performance comparison, \textcolor{black} {when given 10 available UAV-BSs in urban environment parameters, we test 5 algorithms to obtain the coverage percentage of UEs. In each algorithm test, we generated 15 times of uniform UE distributions of 80, 200 and 450 UEs respectively in the same square region. Besides the GA deployment strategy proposed, we have simulated four other schemes for comparison. The first one is random placement which randomly selects a location in a uniform distribution within the square region and a coverage radius. The second one is the K-means algorithm which partitions the UEs into $\hat{K}$ clusters to be covered by $\hat{K}$ UAV-BSs. The third one is a linear programming method called branch and cut~\cite{Botincan2006} which breaks down each UAV-BS placing into sub-problems and optimizes each placement. The fourth one is called greedy search which does the UAV-BS placement one by one and maximizes the covered UEs in each placement. Compared with four other algorithms as shown in Table~\ref{table:2}., namely, K-Means, Branch and Cut, Greedy and Cut, Greedy Search, Random,} GA has demonstrated the significant advantage of solving the optimization problem with many variables involved. It is observed that the result of GA based deployment has higher coverage percentage and this advantage is more pronounced when the number of UEs increases, \textcolor{black}{at the cost of higher complexity and more computing resources.}

\textcolor{black}{Furthermore, the proposed GA algorithm can be potentially applied to more application scenarios, such as, geoscience and remote sensing \cite{Moreira}, cloud computing \cite{Gai}, performing tasks by self-sufficient autonomous robots \cite{Parker}, automatic voltage regulator system design optimization \cite{Hasanien}, etc. Furthermore, a wide range of real-time applications, such as biomedical wireless power transfer (WPT) \cite{Bito}, multi-core systems \cite{Oklapi}, real-time design of thinned array antennas \cite{Cao}, fault repair scheme \cite{Wang_GA}, automatic mode-locked fiber laser \cite{Pu}, traffic surveillance \cite{Lee_2015}, have been realized based on GA algorithms. }

\textcolor{black}{Last, in order to evaluate how the errors of detecting UEs' precise and exact locations affect the numerical results, a simulation is performed in the same area when the UEs' locations have a maximum localization error of 5 meters (a normal localization resolution of outdoor localization techniques, e.g., GPS), in uniform random directions. Consequently, the numerical results of the average coverage ratio are given after running the simulations 15 times, with UEs set to 80, 200 and 450, respectively. From Table~\ref{table:3}, it can be observed that the 5-meter localization error hardly affects the result. Therefore, our proposed method maintains reliable performance even in a practical and challenging application scenario.}

\begin{table}[h]
\vspace{0cm}
\centering
\caption{\textcolor{black}{Comparison of the coverage with and without error in the UEs' locations.}}
\begin{tabular}{||c c c c||} 
 \hline
 \text{Methods--}{$\left | \mathcal P \right | $}  & 80 & 200 & 450 \\ [-0.5ex] 
 \hline\hline
 GA Deployment Method (Without error) & 99.2\% & 88.6\% & 75.3\% \\ 
 GA Deployment Method (With error) & 99.1\% & 88.5\% & 75.1\% \\
 [0ex] 
 \hline
\end{tabular}
\label{table:3}
\end{table}

\section{Conclusions}
\label{sec:conclusion}
This research has proposed and evaluated a cost-efficient 3D UAV-BS deployment algorithm for providing real-life wireless communication services when all the UEs are randomly distributed with various data rate requirements. A novel \textcolor{black}{and practical} GA-based UAVs deployment algorithm has been designed to maximize the number of covered UEs while simultaneously meeting the UEs' individual data rate requirements under the capacity limit of UAV-BSs, \textcolor{black}{which have not been considered in existing works.} The proposed algorithm outperforms four conventional approaches in terms of the coverage ratio, with good tolerance to UEs' localization errors. 

A possible future work is to extend the GA-based deployment algorithm to the applications when A2G interference model is involved. \textcolor{black}{Also, a real experimental validation involving both UAVs and ground users will be interesting to implement to verify the original idea and algorithm proposed in this paper.}   


\ifCLASSOPTIONcaptionsoff
  \newpage
\fi
\bibliographystyle{IEEEtran}
\bibliography{IEEEabrv,biblist}

\begin{thebibliography}{10}
\providecommand{\url}[1]{#1}
\csname url@samestyle\endcsname
\providecommand{\newblock}{\relax}
\providecommand{\bibinfo}[2]{#2}
\providecommand{\BIBentrySTDinterwordspacing}{\spaceskip=0pt\relax}
\providecommand{\BIBentryALTinterwordstretchfactor}{4}
\providecommand{\BIBentryALTinterwordspacing}{\spaceskip=\fontdimen2\font plus
\BIBentryALTinterwordstretchfactor\fontdimen3\font minus
  \fontdimen4\font\relax}
\providecommand{\BIBforeignlanguage}[2]{{%
\expandafter\ifx\csname l@#1\endcsname\relax
\typeout{** WARNING: IEEEtran.bst: No hyphenation pattern has been}%
\typeout{** loaded for the language `#1'. Using the pattern for}%
\typeout{** the default language instead.}%
\else
\language=\csname l@#1\endcsname
\fi
#2}}
\providecommand{\BIBdecl}{\relax}
\BIBdecl

\bibitem{Zeng2016}
Y.~{Zeng}, R.~{Zhang}, and T.~J. {Lim}, ``Wireless communications with unmanned
  aerial vehicles: opportunities and challenges,'' \emph{IEEE Communications
  Magazine}, vol.~54, no.~5, pp. 36--42, May 2016.

\bibitem{LinUAV}
X.~{Lin}, V.~{Yajnanarayana}, S.~D. {Muruganathan}, S.~{Gao}, H.~{Asplund},
  H.~{Maattanen}, M.~{Bergstrom}, S.~{Euler}, and Y.~.~E. {Wang}, ``The sky is
  not the limit: {LTE} for unmanned aerial vehicles,'' \emph{IEEE
  Communications Magazine}, vol.~56, no.~4, pp. 204--210, Apr. 2018.

\bibitem{3GPP}
{3GPP TR 36.777}, ``Technical specification group radio access network: study
  on enhanced {LTE} support for aerial vehicles,'' V15.0.0, Dec. 2017.

\bibitem{Zeng2019a}
\BIBentryALTinterwordspacing
Y.~Zeng, Q.~Wu, and R.~Zhang. Accessing from the sky: A tutorial on {UAV}
  communications for 5{G} and beyond. [Online]. Available:
  \url{http://arxiv.org/abs/1903.05289v2}
\BIBentrySTDinterwordspacing

\bibitem{HuoUAV}
Y.~{Huo}, X.~{Dong}, T.~{Lu}, W.~{Xu}, and M.~{Yuen}, ``Distributed and
  multilayer {UAV} networks for next-generation wireless communication and
  power transfer: A feasibility study,'' \emph{IEEE Internet of Things
  Journal}, vol.~6, no.~4, pp. 7103--7115, Aug. 2019.

\bibitem{HuoMaritime}
Y.~{Huo}, X.~{Dong}, and S.~{Beatty}, ``Cellular communications in ocean waves
  for maritime internet of things,'' \emph{IEEE Internet of Things Journal},
  pp. 1--1, 2020.

\bibitem{DroneFilming}
C.~{Huang}, Z.~{Yang}, Y.~{Kong}, P.~{Chen}, X.~{Yang}, and K.~T. {Cheng},
  ``Through-the-lens drone filming,'' in \emph{2018 IEEE/RSJ International
  Conference on Intelligent Robots and Systems (IROS)}, Oct 2018, pp.
  4692--4699.

\bibitem{EventFilming}
E.~{Natalizio}, N.~{Zema}, E.~{Yanmaz}, L.~{Di Puglia Pugliese}, and
  F.~{Guerriero}, ``Take the field from your smartphone: Leveraging {UAV}s for
  event filming,'' \emph{IEEE Transactions on Mobile Computing}, pp. 1--1,
  2019.

\bibitem{Mademlis2019}
I.~{Mademlis}, V.~{Mygdalis}, N.~{Nikolaidis}, M.~{Montagnuolo}, F.~{Negro},
  A.~{Messina}, and I.~{Pitas}, ``High-level multiple-{UAV} cinematography
  tools for covering outdoor events,'' \emph{IEEE Transactions on
  Broadcasting}, vol.~65, no.~3, pp. 627--635, Sep. 2019.

\bibitem{Lyu2017}
J.~{Lyu}, Y.~{Zeng}, R.~{Zhang}, and T.~J. {Lim}, ``Placement optimization of
  {UAV}-mounted mobile base stations,'' \emph{IEEE Communications Letters},
  vol.~21, no.~3, pp. 604--607, Mar. 2017.

\bibitem{Mozaffari2016}
M.~{Mozaffari}, W.~{Saad}, M.~{Bennis}, and M.~{Debbah}, ``Efficient deployment
  of multiple unmanned aerial vehicles for optimal wireless coverage,''
  \emph{IEEE Communications Letters}, vol.~20, no.~8, pp. 1647--1650, Aug.
  2016.

\bibitem{Galkin2016}
B.~{Galkin}, J.~{Kibilda}, and L.~A. {DaSilva}, ``Deployment of {UAV}-mounted
  access points according to spatial user locations in two-tier cellular
  networks,'' in \emph{Proc. Wireless Days (WD)}, Mar. 2016, pp. 1--6.

\bibitem{Al-Hourani2014}
A.~{Al-Hourani}, S.~{Kandeepan}, and S.~{Lardner}, ``Optimal lap altitude for
  maximum coverage,'' \emph{IEEE Wireless Communications Letters}, vol.~3,
  no.~6, pp. 569--572, Dec. 2014.

\bibitem{Bor-Yaliniz2016}
R.~I. {Bor-Yaliniz}, A.~{El-Keyi}, and H.~{Yanikomeroglu}, ``Efficient {3-D}
  placement of an aerial base station in next generation cellular networks,''
  in \emph{Proc. IEEE Int. Conf. Communications (ICC)}, May 2016, pp. 1--5.

\bibitem{Alzenad2017}
M.~{Alzenad}, A.~{El-Keyi}, F.~{Lagum}, and H.~{Yanikomeroglu}, ``{3-D}
  placement of an unmanned aerial vehicle base station ({UAV-BS}) for
  energy-efficient maximal coverage,'' \emph{IEEE Wireless Communications
  Letters}, vol.~6, no.~4, pp. 434--437, Aug. 2017.

\bibitem{Hou-2019}
T.~{Hou}, Y.~{Liu}, X.~{Sun}, Z.~{Song}, and Y.~{Chen}, ``Non-orthogonal
  multiple access in multi-{UAV} networks,'' in \emph{2019 IEEE 90th Vehicular
  Technology Conference (VTC2019-Fall)}, Sep. 2019, pp. 1--5.

\bibitem{Feng-2019}
A.~A. {Khuwaja}, G.~{Zheng}, Y.~{Chen}, and W.~{Feng}, ``Optimum deployment of
  multiple {UAV}s for coverage area maximization in the presence of co-channel
  interference,'' \emph{IEEE Access}, vol.~7, pp. 85\,203--85\,212, 2019.

\bibitem{Liu-2018}
L.~{Liu}, S.~{Zhang}, and R.~{Zhang}, ``Cooperative interference cancellation
  for multi-beam {UAV} uplink communication: A {D}o{F} analysis,'' in
  \emph{2018 IEEE Globecom Workshops (GC Wkshps)}, Dec 2018, pp. 1--6.

\bibitem{Huo-VTC-2019}
Y.~{Huo}, F.~{Lu}, F.~{Wu}, and X.~{Dong}, ``Multi-beam multi-stream
  communications for {5G} and beyond mobile user equipment and {UAV} proof of
  concept designs,'' in \emph{2019 IEEE 90th Vehicular Technology Conference
  (VTC2019-Fall)}, Sep. 2019, pp. 1--5.

\bibitem{UAV-System}
D.~{Hu}, Q.~{Zhang}, Q.~{Li}, and J.~{Qin}, ``Joint position, decoding order,
  and power allocation optimization in {UAV}-based {NOMA} downlink
  communications,'' \emph{IEEE Systems Journal}, pp. 1--12, 2019.

\bibitem{Mei-2019}
W.~{Mei} and R.~{Zhang}, ``Uplink cooperative noma for cellular-connected
  uav,'' \emph{IEEE Journal of Selected Topics in Signal Processing}, vol.~13,
  no.~3, pp. 644--656, June 2019.

\bibitem{Liu-2019}
L.~{Liu}, S.~{Zhang}, and R.~{Zhang}, ``Multi-beam {UAV} communication in
  cellular uplink: Cooperative interference cancellation and sum-rate
  maximization,'' \emph{IEEE Transactions on Wireless Communications}, vol.~18,
  no.~10, pp. 4679--4691, Oct 2019.

\bibitem{Al-Turjman2019}
F.~{Al-Turjman}, J.~P. {Lemayian}, S.~{Alturjman}, and L.~{Mostarda},
  ``Enhanced deployment strategy for the {5G} drone-{BS} using artificial
  intelligence,'' \emph{IEEE Access}, vol.~7, pp. 75\,999--76\,008, 2019.

\bibitem{Slowik2018}
A.~{Slowik} and H.~{Kwasnicka}, ``Nature inspired methods and their industry
  applications—swarm intelligence algorithms,'' \emph{IEEE Transactions on
  Industrial Informatics}, vol.~14, no.~3, pp. 1004--1015, 2018.

\bibitem{Govindan}
K.~{Govindan}, A.~{Jafarian}, and V.~{Nourbakhsh}, ``Designing a sustainable
  supply chain network integrated with vehicle routing: A comparison of hybrid
  swarm intelligence metaheuristics,'' \emph{Computers \& Operations Research},
  2015.

\bibitem{Zhao_2019}
X.~{Zhao}, C.~{Wang}, J.~{Su}, and J.~{Wang}, ``Research and application based
  on the swarm intelligence algorithm and artificial intelligence for wind farm
  decision system,'' \emph{Renewable Energy}, vol. 134, pp. 681--697, 2019.

\bibitem{Brezocnik}
L.~{Brezo{\v c}nik}, I.~{Fister}, and V.~{Podgorelec}, ``Swarm intelligence
  algorithms for feature selection: a review,'' \emph{Applied Sciences},
  vol.~8, no.~9, p. 1521, Sep. 2018.

\bibitem{Anandakumar}
H.~{Anandakumar} and K.~{Umamaheswari}, ``A bio-inspired swarm intelligence
  technique for social aware cognitive radio handovers,'' \emph{Computers \&
  Electrical Engineering}, vol.~71, pp. 925--937, 2018.

\bibitem{Dulebenets}
M.~A. {Dulebenets}, ``A comprehensive evaluation of weak and strong mutation
  mechanisms in evolutionary algorithms for truck scheduling at cross-docking
  terminals,'' \emph{IEEE Access}, vol.~6, pp. 65\,635--65\,650, 2018.

\bibitem{Dulebenets_2019}
------, ``A delayed start parallel evolutionary algorithm for just-in-time
  truck scheduling at a cross-docking facility,'' \emph{International Journal
  of Production Economics}, vol. 212, pp. 236--258, 2019.

\bibitem{Vahdani}
B.~{Vahdani} and S.~{Shahramfard}, ``A truck scheduling problem at a
  cross-docking facility with mixed service mode dock doors,''
  \emph{Engineering Computations.}, vol.~36, pp. 1977--2009, 2019.

\bibitem{Yang2020}
Y.~{Yang} and C.~{Yang}, ``Research on the application of {GA} improved neural
  network in the prediction of financial crisis,'' in \emph{Proc. 12th Int.
  Conf. Measuring Technology and Mechatronics Automation (ICMTMA)}, 2020, pp.
  625--629.

\bibitem{GA-BS-2019}
Y.~{Liu}, W.~{Huangfu}, H.~{Zhang}, H.~{Wang}, W.~{An}, and K.~{Long}, ``An
  efficient geometry-induced genetic algorithm for base station placement in
  cellular networks,'' \emph{IEEE Access}, vol.~7, pp. 108\,604--108\,616,
  2019.

\bibitem{Goldberg:1989:GAS:534133}
D.~E. Goldberg, \emph{Genetic Algorithms in Search, Optimization and Machine
  Learning}, 1st~ed.\hskip 1em plus 0.5em minus 0.4em\relax Boston, MA, USA:
  Addison-Wesley Longman Publishing Co., Inc., 1989.

\bibitem{Herrera2001}
F.~{Herrera} and M.~{Lozano}, ``Adaptive genetic operators based on coevolution
  with fuzzy behaviors,'' \emph{IEEE Transactions on Evolutionary Computation},
  vol.~5, no.~2, pp. 149--165, 2001.

\bibitem{Botincan2006}
M.~{Botincan} and G.~{Nogo}, ``Anomalies in distributed branch-and-cut solving
  of the capacitated vehicle routing problem,'' in \emph{Proc. 28th Int. Conf.
  Information Technology Interfaces}, 2006, pp. 677--682.

\bibitem{Moreira}
L.~P. {Moreira}, ``Time-domain receiver function deconvolution using genetic
  algorithm,'' \emph{IEEE Geoscience and Remote Sensing Letters}, pp. 1--5,
  2019.

\bibitem{Gai}
K.~{Gai}, M.~{Qiu}, and H.~{Zhao}, ``Cost-aware multimedia data allocation for
  heterogeneous memory using genetic algorithm in cloud computing,'' \emph{IEEE
  Transactions on Cloud Computing}, pp. 1--1, 2016.

\bibitem{Parker}
G.~{Parker} and R.~{Zbeda}, ``Learning area coverage for a self-sufficient
  hexapod robot using a cyclic {Genetic} {Algorithm},'' \emph{IEEE Systems
  Journal}, vol.~8, no.~3, pp. 778--790, 2014.

\bibitem{Hasanien}
H.~M. {Hasanien}, ``Design optimization of {PID} controller in automatic
  voltage regulator system using {Taguchi} combined genetic algorithm method,''
  \emph{IEEE Systems Journal}, vol.~7, no.~4, pp. 825--831, 2013.

\bibitem{Bito}
J.~{Bito}, S.~{Jeong}, and M.~M. {Tentzeris}, ``A real-time electrically
  controlled active matching circuit utilizing genetic algorithms for
  biomedical wpt applications,'' in \emph{2015 IEEE Wireless Power Transfer
  Conference (WPTC)}, 2015, pp. 1--4.

\bibitem{Oklapi}
E.~{Oklapi}, M.~{Deubzer}, S.~{Schmidhuber}, E.~{Lalo}, and J.~{Mottok},
  ``Optimization of real-time multicore systems reached by a {Genetic}
  {Algorithm} approach for runnable sequencing,'' in \emph{2014 International
  Conference on Applied Electronics}, 2014, pp. 233--238.

\bibitem{Cao}
D.~{Cao}, A.~{Modiri}, G.~{Sureka}, and K.~{Kiasaleh}, ``{DSP} implementation
  of the particle swarm and {Genetic Algorithms} for real-time design of
  thinned array antennas,'' \emph{IEEE Antennas and Wireless Propagation
  Letters}, vol.~11, pp. 1170--1173, 2012.

\bibitem{Wang_GA}
J.~{Wang}, J.~{Kang}, and G.~{Hou}, ``Real-time fault repair scheme based on
  improved {Genetic Algorithm},'' \emph{IEEE Access}, vol.~7, pp.
  35\,805--35\,815, 2019.

\bibitem{Pu}
G.~{Pu}, L.~{Yi}, L.~{Zhang}, and W.~{Hu}, ``{Genetic Algorithm}-based fast
  real-time automatic mode-locked fiber laser,'' \emph{IEEE Photonics
  Technology Letters}, vol.~32, no.~1, pp. 7--10, 2020.

\bibitem{Lee_2015}
G.~{Lee}, R.~{Mallipeddi}, G.~{Jang}, and M.~{Lee}, ``A {Genetic
  Algorithm}-based moving object detection for real-time traffic
  surveillance,'' \emph{IEEE Signal Processing Letters}, vol.~22, no.~10, pp.
  1619--1622, 2015.

\end{thebibliography}

\begin{IEEEbiography}[{\includegraphics[width=1in,height=1.25in,keepaspectratio]{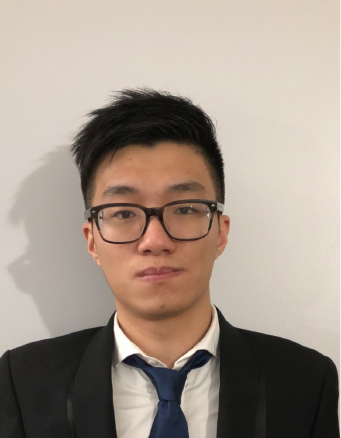}}]{Xukai Zhong}received his Bachelor degree of Applied  Science from Simon Fraser University, Burnaby, BC, Canada, in 2018, and Master degree of Engineering in Electrical and Computer Engineering from University of Victoria, Victoria, BC, Canada, in 2020. He is currently working in Fortinet, Burnaby, BC, Canada, as Software Developing Engineer. His recent research interests include machine learning, computer vision, robotics and UAV communications. 
\end{IEEEbiography}

\begin{IEEEbiography}[{\includegraphics[width=1in,height=1.25in,keepaspectratio]{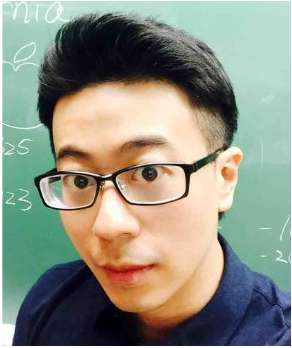}}]{Yiming Huo}(S'08--M'18) received his B.Eng degree in information engineering from Southeast University, China, in 2006, and M.Sc. degree in System-on-Chip (SoC) from Lund University, Sweden, in 2010, and Ph.D. in electrical engineering at University of Victoria, Canada, in 2017, and he is currently a Post-Doctoral Research Fellow with the same department. His recent research interests include 5G and beyond wireless systems, terahertz technology, space technology, Internet of Things, and machine learning. 

He has worked in several companies and institute including Ericsson, ST-Ericsson, Chinese Academy of Sciences, STMicroelectronics, and Apple Inc., Cupertino, CA, USA. He is a member of several IEEE societies, and also a member of the Massive MIMO Working Group of the IEEE Beyond 5G Roadmap. He was a recipient of the Best Student Paper Award of the 2016 IEEE ICUWB, the Excellent Student Paper Award of the 2014 IEEE ICSICT, and the Bronze Leaf Certificate of the 2010 IEEE PrimeAsia. He also received the ISSCC-STGA Award from the IEEE Solid-State Circuits Society (SSCS), in 2017. He has served as the Program Committee of the IEEE ICUWB 2017, the TPC of the IEEE VTC 2018/2019/2020, the IEEE ICC 2019, the Session Chair of the IEEE 5G World Forum 2018, the Publication Chair of the IEEE PACRIM 2019, the Technical Reviewer for multiple premier IEEE conferences and journals. He is currently an Associate Editor for \emph{IEEE Access}.

\end{IEEEbiography}

\begin{IEEEbiography}[{\includegraphics[width=1in,height=1.25in,keepaspectratio]{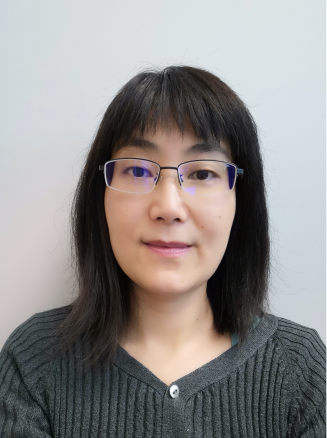}}]{Xiaodai Dong}(S'97--M'00--SM'09) received the B.Sc. degree in information and control engineering from Xi’an Jiaotong University, China, in 1992, the M.Sc. degree in electrical engineering from the National University of Singapore in 1995, and the Ph.D. degree in electrical and computer engineering from Queen’s University, Kingston, ON, Canada, in 2000. From 1999 to 2002, she was with Nortel Networks, Ottawa, ON, Canada, and worked on the base transceiver design of the third-generation mobile communication systems. From 2002 to 2004, she was an Assistant Professor with the Department of Electrical and Computer Engineering, University of Alberta, Edmonton, AB, Canada. Since 2005, she has been with the University of Victoria, Victoria, Canada, where she is currently a Professor with the Department of Electrical and Computer Engineering. She was the Canada Research Chair (Tier II) from 2005 to 2015. 

Dr. Dong's research interests include 5G, mmWave communications, radio propagation, Internet of Things, machine learning, terahertz communications, localization, wireless security, e-health, smart grid, and nano-communications. She served as an Editor for \emph{IEEE Transactions on Wireless Communications} in 2009-2014, \emph{IEEE Transactions on Communications} in 2001-2007, \emph{Journal of Communications and Networks} in 2006-2015, and is currently an Editor for \emph{IEEE Transactions on Vehicular Technology} and \emph{IEEE Open Journal of the Communications Society}. 

\end{IEEEbiography}

\begin{IEEEbiography}[{\includegraphics[width=1in,height=1.25in,keepaspectratio]{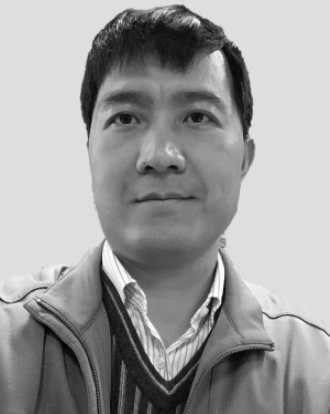}}]{Zhonghua Liang} (S'07--M'08--SM'20) received the B.Sc. degree in radio engineering and the M.Sc. and Ph.D. degrees in information and communication engineering from Xi'an Jiaotong University, Xi'an, China, in 1996, 2002, and 2007, respectively. From July 1996 to August 1999, he was with the Guilin Institute of Optical Communications (GIOC), Guilin, China, where he was a System Engineer in optical transmission systems. From 2008 to 2009, he was a Postdoctoral Fellow with the Department of Electrical and Computer Engineering, University of Victoria, Victoria, BC, Canada. Since 2010, he has been with the School of Information Engineering, Chang'an University, Xi'an, China, where he is currently a Professor. His research interests include ultra-wideband technology, wireless communication theory, Internet of Things, wireless sensor networks, and adaptive signal processing techniques for wireless communication systems.

\end{IEEEbiography}

\end{document}